\documentstyle[12pt]{article}
\newcommand{\hs}{\hspace}
\newcommand{\vs}{\vspace}
\newcommand{\beq}{\begin{equation}}
\newcommand{\eeq}{\end{equation}}
\newcommand{\beqa}{\begin{eqnarray}}
\newcommand{\eeqa}{\end{eqnarray}}
\newcommand{\beqan}{\begin{eqnarray*}}
\newcommand{\eeqan}{\end{eqnarray*}}

\newcommand{\ol}{\overline}
\newcommand{\ra}{\rightarrow}

\newcommand{\ben}{\begin{enumerate}}
\newcommand{\een}{\end{enumerate}}
\newcommand{\bfl}{\begin{flushleft}}
\newcommand{\efl}{\end{flushleft}}
\newcommand{\ba}{\begin{array}}
\newcommand{\ea}{\end{array}}
\newcommand{\btab}{\begin{tabular}}
\newcommand{\etab}{\end{tabular}}
\newcommand{\bit}{\begin{itemize}}
\newcommand{\eit}{\end{itemize}}

\begin{document}

 \pagestyle{plain}
 \thispagestyle{empty}

 \vspace*{-10mm}

 \begin{flushright}
 {\bf BROWN-HET-941} \\
 {\bf BROWN-TA-506} \\
 {\bf hep-ph/9404351} \\
 {\bf April 1994} \\
 \end{flushright}

 \vspace{1.0cm}
 \begin{center}
 {\Large \bf High Energy Hadron-Hadron Scattering
 \footnote{Presented at the 6th Summer School \& Symposium on Nuclear
 Physics, 20-25 August 1993 (Moojoo, Korea),
 and supported in part by the
 US Department of Energy Contract DE-FG02-91ER40688-Task A
 and also in part by the Brown-SNUCTP Exchange Program.} } \\
 \vglue 15mm
 {\bf Kyungsik Kang } \\
 \vglue 2mm
 \footnotesize
 {\it Department of Physics, Brown University
 , Providence, RI 02912, USA\footnote{Permanent Address}, } \\
 \vglue 1mm
 {\it Center for Theoretical Physics, Seoul National University,
 Seoul 151-742, Korea, } \\
 \vglue 1mm
 {\it and} \\
 \vglue 1mm
 {\it Division de Physique Th\'eorique, Institut de Physique Nucl\'eaire,
91406 Orsay Cedex and } \\
 {\it LPTPE, Universit\'e P. \& M. Curie, 4 Place Jussieu, 75252 Paris,
Cedex 05, France } \\
 \vglue 10mm
 \normalsize
 {\bf ABSTRACT} \\
 \vglue 5mm
 \begin{minipage}{14cm}
 {\normalsize
 The elastic hadron-hadron scattering at high energies is
 one of the most
 fundamental subjects of all particle
 physics
 problems and yet is
 least understood in spite of many advances in
 quantum chromodynamics (QCD)
 at
 the conceptual level. We review here the
 recent theoretical and
 experimental
status of the subject as well as
 the rigorous results
 of the high energy hadron-hadron scattering.
 Surprisingly enough,
 the high-energy models for the elastic and
 diffractive scattering
 abstracted from the Regge-Pomeron field theory
 is still
 {\it phenomenologically} successful to explain the
 high-energy scattering
 data though no rigorous derivation out of the QCD
 is yet available.
 }
 \end{minipage}
 \end{center}
 %\vglue 20mm
 %%%%%%%%%%%%%%%%%%%%%%%%%%%%%%%%%%%%%%%%%%%%%%%%%%%%%%%%%%%%%%%%%%%%%%%%%
 \newpage
 %%%%%%%%%%%%%%%%%%%%%%%%%%%%%%%%%%%%%%%%%%%%%%%%%%%%%%%%%%%%%%%%%%%%%%%%%
 %
 \section{Introduction}
 With the recent report from the UA4/2 group [1] on new determination of
 $\rho $ for the forward $p\bar p$ scattering amplitude at CERN
 $Sp\bar{p}S$ energy $\sqrt{s}$ = 541 GeV which replaced the earlier
 high $\rho $ value [2] and with the new measurements of the total
 cross-section $\sigma_T$ by the CDF group [3] and the E710 group [4] at
 Fermilab Tevatron
 energy $\sqrt{s}$ = 1.8 TeV, it seems appropriate and necessary to
 review the status of the high energy hadron-hadron scattering and
 put our current understanding of the high-energy strong interactions
 into perspective. Indeed these are basically the complete information
 that we will have on the high-energy elastic scattering until the
 LHC is built and becomes operative.

 Though we have in principle the exact theory of the strong interactions,
 QCD, which can give good descriptions of the hadron interactions
 at short distances, i.e. the high-energy deep-inelastic reactions,
 the interactions at large distances, namely, near forward scattering
 can not reliably be calculated in the framework of QCD. However
 the two domains are intimately related to each other and the description
 of the high-energy elastic and diffractive scattering based on
 the microscopic gauge theory of colored quarks and gluons, QCD, presents
 a major challenge.
 Indeed it is a central problem of the modern theory
 of strong interactions to establish
 a
 smooth connection and
 a
 relationship
 between the {\it hard Pomeron} obtained from perturbative QCD
 for the hard processes and the {\it soft Pomeron} emerged from
 phenomenological models for the high-energy near forward hadron-hadron
 scattering[5].
 The discovery of the {\it semihard} processes in QCD for the
 high-energy inclusive interactions may eventually lead to
 matching of the two regions but
 the gap between the phenomenological models
 for the high-energy hadron scattering and the microscopic
 gauge theory of the
 strong interaction is so wide for the moment that it is
 difficult to imagine any relationship between them.

 Another view is that the strong interactions at large distances can
 at least be approximately calculated by lattice QCD. But the continuum
 limit of the realistic lattice QCD, exhibiting the color confining
 and unitary hadron interactions and asymptotically free interactions
 at short distances, seems to be as remote as the asymptotic energy
 region, {\it asymptopia}, of the hadron interactions.

 The concept of the Pomeron was originally introduced to understand
 the high-energy behavior of total cross sections
 which is controlled by a right-most Regge pole with vacuum quantum number
 in the analytic complex angular momentum plane.
 With the advent of the analytic S-matrix consistent with the
 requirement of the s-channel unitarity, the dual resonance model was invented,
 which can be identified as
 the tree-level amplitudes in the string theory, thus raising a
 hope to find the theory of the strong interactions in the string
 theory approach for the colored quarks and gluons that may lead to
 the Reggeon calculus based on the microscopic structure
 of QCD. But the
 search for the realistic string theory with the desired QCD
 properties in both infrared and
 ultraviolet
 regions has not
 yet been successful. It is however clear that in any
 of these developments the Pomeron plays the central role
 in describing the high-energy analytic S-matrix consistent
 with the unitarity requirement.

 In the Reggeon calculus that was invented before the gauge
 theory of QCD, it is well-known that one must include
 the multi-Pomeron exchanges to consistently describe the
 high energy analytic S-matrix because of the unitarity.
 This means that the single {\it bare Pomeron} has to be
 renormalized by rescattering of two Pomerons via
 multi-Pomeron exchanges. It turns out that
 this procedure can be compared to the eikonal formalism which
 guarantees the manifest s-channel unitarity and converts the bare Pomeron
 power increase of total cross sections to a logarithmic increase
 of the renormalized Pomeron consistent with the rigorous unitarity limit.
 Constructing a microscopic theory of colored quarks and gluons
 that can give the bare Pomeron intercept slightly larger than one
 is then a major task of the Reggeon calculus. Indeed it is well-known
 that the ladder diagram for scattering of two reggeized gluons in
 the t-channel in the leading-log approximation can generate a fixed
 branch-point
 singularity
 in the angular momentum plane, i.e.,
 the Lipatov Pomeron [6], at ${\it j } = 1 + \epsilon $ with
 $\epsilon = \frac{g^2}{\pi^2} 2 \ln 2$ so that $\sigma_T \sim s^\epsilon$
 which violates the Froissart bound. When the perturbative leading-log
 QCD is extended to the exchange of N-gluons, it produces a sequence of
 branch points on an interval $j = 1 + \epsilon_N $ where
 $\epsilon_N \le \frac{1}{2} N(N-1)
 \epsilon $. Thus it seems very likely that
 such a dense sequence of singularities gives such a
 complicated Pomeron structure that it may not
 have a chance to reconcile the s-channel unitarity. In addition,
 the calculations leading to $\epsilon $ and $\epsilon_N$ produce
 amplitudes [7] which have zero transverse momentum singularities
 (but not divergences) due to the exchange of massless gluons
 thus conflicting
 with
 the color confinement and absence of zero mass
 particles in the strong interactions. The discovery of the semihard
 processes, i.e., the production of quark- and gluon-jets with
 relatively large transverse momenta but with a small fraction
 of the initial energy may help to get around some of the difficulties
 for the leading-log approximation of perturbative QCD [8] but
 it appears one needs additional phenomenological inputs
 such as the structure functions of the partons
 in the initial hadrons and the fragmentation functions of the
 quarks and gluons into the secondary hadrons. In addition, a
 smooth matching of the semihard Pomeron to the soft Pomeron
 responsible for large cross sections of the elastic and
 diffractive scattering has not yet been explicitly
 demonstrated for the realistic color interactions.

 There is
 another
 somewhat pessimistic view that matching the low
 $p_T$ physics and the ultraviolet QCD of parton models and perturbation
 theory would require a new technology [9], whereby proposing a program
 to extract the low $p_T$ Pomeron from QCD which is based on an
 analysis of the infrared divergences (as the gauge symmetry
 is restored) of a reggeon diagram description of spontaneously
 broken QCD. Needless to say, the description of the large cross
 section physics at high energies in terms of the microscopic
 structure of QCD is far from the reality and one must still rely
 on the phenomenological models that satisfy the general principles
 required for the analytic S-matrix and
 utilize
 the general
 QCD concepts only.

 In building phenomenological models for the high-energy scattering,
 asymptotic analytic amplitude analysis based on
 the general principles of unitarity, analyticity and crossing-symmetry
 has proven to be useful even though the currently
 accessible
 energy region is not in asymptopia.
 In particular, asymptotic amplitude analysis led to the quasilocal
 analyticity relations expressed in terms of  derivative dispersion
 relations (DDR) with respect to $\ln s$ variable at energies
 sufficiently higher than the typical resonance domain for both even-
 and odd-signatured amplitudes [10]. Phenomenological models can then be
 constructed as the solution of DDR that
 satisfy
 the unitarity bound,
 and have proven to be successful in understanding the high-energy
 scattering data, though not asymptotic, as well as predicting
 the high-energy behavior of the hadron-hadron scattering amplitude [11].
 The total cross sections for $pp$ and $p\bar{p}$ scattering
 do indeed smoothly rise in the ISR
 energies, suggesting {\it a priori} justification of the use
 of the quasilocal analyticity relation at the current high energies.
 Also there have been comprehensive eikonal amplitude analyses [12]
 based on QFD expectations and motivations that {\it mimic} QCD. Though
 different in derivations, they both can give the same asymptotic term
 responsible for the characteristic rise of the total cross section
 at high energies: the asymptotically increasing term in the
 analytic amplitude
 analysis is due to a leading $j-$plane singularity near the forward
 direction constrained by and consistent with rigorous asymptotic
 theorems, while such a
 term can also be constructed out of an
 eikonal that is motivated by general field theory expectations.
 The asymptotic term in the amplitude has a
 natural interpretation
 as coming from the leading $j-$plane singularity, i.e., that has
 the intercept $1$ in the forward direction.
 Within the general framework of QCD, the asymptotic term in the
 crossing-even amplitude $F_{+}(s,t)$ can be thought of as the
 renormalized or physical Pomeron that is achieved by unitarization
 of the bare Pomeron of color-singlet and C-even two-gluon exchanges
 through rescattering of two gluons to include multigluon exchanges.
 If the leading bare $j-$plane singularity due to the C-odd and
 color-singlet three-gluon exchange has the intercept $\alpha_{o}(0)
 = 1 + \epsilon_{o} > 1 $, this will imply a nonvanishing odd-signatured
 amplitute $F_{-} (s, t=0)$ as $s {\to} {\infty}$. This C-odd
 leading $j-$plane singularity is the C-odd counterpart of the Pomeron
 and became to be known as the {\it Odderon.}

 One can easily introduce both the Odderon and Pomeron in the
 asymptotic amplitude
 analysis as solutions of the DDR [10] that
 maximally saturate the rigorous bounds of the amplitudes
 for both
 particle-particle and antiparticle-particle scattering so that
 both $F_{\pm}$ can grow as fast as maximally possible at asymptotic
 energies. The maximal Odderon model seems to have no
 support from the near forward scattering experiments by now,
 although it can arise naturally as
 a reggeized three-gluon exchange
 in the leading-log approximation of perturbative
 QCD. It is however not clear if its intercept is indeed larger
 than 1. In addition, even if the bare Odderon intercept is
 assumed to be bigger than one, it is surprisingly difficult
 to generate large odd-signatured amplitude within the eikonal
 formalism as it is incapable of giving the maximal Odderon behavior
 [13]. Though the Odderon is irrelevant for the forward scattering,
 one can not rule out its role in the non-forward region,
 in particular if there is a
 non-zero difference between the
 differential cross sections of $pp$ and ${\bar p}p$.
 It has been suggested [14] that the three-gluon exchange
 in the form of a Regge cut might be important
 in order to understand the energy dependence of the diffractive minimum
 and secondary maximum
 region of the differential cross sections through the
 interference with the Pomeron and Pomeron-Pomeron cut.
 The new data for $\rho$ from UA4/2 has washed out any need
 to include the Odderon contribution to the high-energy forward
 elastic scattering.

 There are a number of rigorous results for the analytic
 S-matrix at high energies that are based on some
 very general principles such as unitarity, analyticity
 and crossing-symmetry. Strictly speaking, these results
 are valid in asymptotic energy limit, i.e., the asymptotic
 theorems. Nevertheless, they are useful to guide the
 phenomenological models for the high-energy hadron-hadron
 scattering, in either the asymptotic analytic amplitude
 analysis or the eikonal types mimicking
 general QCD
 properties, and predicting the asymptotic
 form of the rising cross sections,
 which has direct implications on the structure of
 the physical Pomeron. We will see, however, from the
 asymptotic amplitude analysis that the currently
 available high-energy data do not necessarily
 prefer $\ln ^2{s}$ rise over
 $\ln{s}$ behavior of the total cross sections
 in terms of the statistics, despite many common beliefs.
 In fact, the asymptotic theorems have often been
 misinterpreted and even misused in some cases.
 For these
 reasons we will summarize some of these theorems
 in Section III as soon as we establish the notations,
 definitions, and concepts of the hadron-hadron scattering
 at high energies in Section II, in which general features
 of the high-energy scattering data will also be reviewed.
 We then will review in Section IV the status of the
 large cross section processes, namely, the high-energy
 near forward scattering based on the asymptotic amplitude
 analysis and give predictions of the total cross section
 at the LHC energy. We shall be mostly discussing the
 case of the two equal mass particle scattering, namely
 $pp$ and $p\bar p$ scattering amplitudes, for which there
 are data available up to the ISR energies. In addition
 we have high-energy data for $p\bar p$ at the CERN S$p\bar p$
 S and Fermilab Tevatron energies, which provide a raison
 d'\^etre
 for reviewing the features and status
 of the high-energy hadron-hadron scattering. Even though
 the elastic scattering is one of the simplest subjects,
 we will see that it is one of the least understood
 problem from the basic theory of QCD and also still poses
 problems for experiments to clarify.

 \section{Scattering Amplitudes, Cross Sections and Experimental Data}

 Following the notations of Ref. [11], we use the normalization
 convention
 for the scattering amplitude such that the imaginary
 part of the forward elastic scattering amplitude $F (s,t=0)$
 is related to the total cross section $\sigma_{T}$
 by the optical theorem as
\begin{equation}
 \sigma_T = \frac{1}{s} \ Im \hs{1mm}F(s,t=0) \label{eq1}
\end{equation}
 for both $pp$ and $p\bar p$ scattering. The crossing-even and
 -odd amplitudes $F_{\pm}$ are given by
\begin{equation}
 F_{\pm} = \frac{1}{2} \ (F_{pp} \pm F_{{\bar p} p})
\end{equation}
 in terms of the $pp$ and ${\bar p}p$ scattering amplitudes,
 so that
\begin{equation}
 F_{pp} = F_{+} + F_{-}  \:,\: F_{\bar {p}p} = F_{+} - F_{-} .
\end{equation}
 In addition
 to the total cross section, experiments measure the
 ratio of the real part to the imaginary part of the forward
 elastic amplitude,
\begin{equation}
 \rho_{\it i} (s) = \frac{Re \hs{1mm} F_{\it i} (s,t=0) }{Im \hs{1mm} F_{\it i}
 (s,t=0)} \ ,  \: ({\it i} = pp, \, p{\bar p}).
\end{equation}
 The differential cross sections for pp and $p\bar{p}$ scattering
 get the Coulomb contribution [15] in addition to the hadronic
 amplitudes,
\begin{equation}
 \frac{d\sigma }{dt} \ = \frac{1}{16 \pi s^2 (\hbar c)^2} \ \vert F_{C}
 + F\vert^2,
\end{equation}
 where
\begin{eqnarray}
 F_{C} = \frac{8 \pi s \alpha (\hbar c)^2 }{\vert t\vert} \ G^2 (t)
 \exp ^{-{\it i}{\alpha}{\phi (t)}}, \\
 \phi (t) = \ln (\frac{0.08}{\vert t\vert} \ ) - 0.577,
\end{eqnarray}
 G(t) being the electromagnetic form factor of the proton
 which is usually parametrized as a dipole form,
 $G(t) = (1 + \frac{\vert t\vert}{0.71} \ )^{-2}$.
 Here $t$ is in units of GeV$^2$. For small $t$ within the
 diffractive region, experimentalists use a simple
 optical model form with a single exponential term
 for the hadronic amplitudes,
\begin{equation}
 F(s,t) = s \hs{1mm} {\sigma_T (s)} \hs{1mm} ({\rho (s)} + {\it i}) \hs{1mm}
 \exp^{-\frac{1}{2} \ B \vert t\vert} \label{eq8}
\end{equation}
 where $B$ is the nuclear slope parameter, though
  it depends on $s$ and $t$ strictly speaking.
  In principle $\sigma_T$, $\rho $, and $B$ are
  the three quantities that
  can be determined from the experimental data of
  differential cross sections at a given energy.
  This is the case in the E710 experiment.
  In addition, $\sigma_T$ can be obtained from
  the simultaneous measurements of the elastic
  differential cross section at small $t$ and the
  total inelastic rates, i.e., the luminosity
  independent method [16], which was used in the ISR,
  UA4, and CDF experiments. Basically the luminosity
 independent method is the only way
  to determine the Tevatron luminosity
  and the total cross section at Fermilab [3].
This method is based on the observation that the total cross section is given
by the sum of the total elastic and inelastic rates,
\beq
\sigma_T = \frac{1}{L} \ ( R_e + R_i ),
\eeq
while the differential cross section is related to the differential  rate
with respect to $t$,
\beq
\frac{d\sigma}{dt} \ = \frac{1}{L} \ \frac{dR_e}{dt}.\
\eeq
Because of (5), we then get
\beq
(1 + \rho^2) \sigma_T = 16 \pi (\hbar c)^2 \hs{1mm} [ \frac{dR_e}{dt} \ ]_{t =
 0}
\hs{1 mm} / \hs{1mm} ( R_e + R_i ).
\eeq
Using this method the CDF collaboration recently obtained
$(1 + \rho^2) \hs{1mm} \sigma_T(s) = 62.64 \pm 0.95 \hspace{2mm}mb $ and $81.83
 \pm 2.29 \hspace{2mm}mb $ at $\sqrt{s} = 54.6 \hspace{2mm} GeV $ and $1800
 \hspace{2mm} GeV $ respectively which agrees with the UA4 value reported
 earlier $63.3 \pm 1.5 \hspace{2mm} mb $ at the CERN $Sp\overline{p}S $ energy
 $\sqrt{s} = 546 \hspace{2mm} GeV $ but disagrees somewhat with the E710 value
 $73.6 \pm 3.3 \hspace{2mm} mb $ based on the three parameter fits to  the
 t-distribution of the differential rates at the Fermilab Tevatron energy. The
 new UA4/2 $\rho $-value is extracted from the $t$-distribution of the
 differential rate with the UA4 value of $(1 + \rho^2) \sigma_T$ as input.
The differential cross section exhibit a sharp Coulomb peak near $\vert t \vert
 $ = 0 and an interference structure around  $\vert t \vert  = 10^{-3}
 \hspace{2mm} GeV^2$ followed by a diffractive pattern of the hadronic
reactions
 which has a characteristic $e^{- B \vert t \vert}$ decrease to a minimum
 followed by a shoulder.
In order to extract the hadronic scattering parameters, the Coulomb
contribution
 has to be carefully accounted for by studying the interference region and
 beyond.
The minimum $\vert t\vert $ achieved by both the new E710 and UA4/2 experiments
 is $0.75 \times 10^{-3} \hspace{2mm} GeV^2 $ well inside the Coulomb-hadron
 interference region.

Fortunately the Coulomb contributions decrease rather rapidly as $\vert t
\vert$
 increases, e.g., it is only about 1 \% at $\vert t \vert = 0.025 \hspace{2mm}
 GeV^2$ the minimum value in the CDF experiment.
The region of maximum interference is expected from where $F_C \sim F$, i.e.,
 $\vert t \vert \sim 8 \pi \alpha (\hbar c)^2 / \sigma_T $, which gives $\vert
t
 \vert \sim 1.1 \times 10^{-3}\hspace{2mm} GeV^2$ for $\sigma_T = 60
 \hspace{2mm}mb$, a reasonable value of the total cross section around
$\sqrt{s}
 = 540 \hspace{2mm} GeV $.
We give in Table 1 basically the complete information of the high-energy
forward
 scattering parameters that are available now.

\vspace{3mm}
\begin{center} \small
\begin{tabular}{|c||c|c|c|c|c|} \hline
 & $\sqrt{s} (GeV)$ & (1+$\rho^2)\sigma_T(mb)$ & $\sigma_T$ & $\rho$ & \it B
 $(GeV)^{-2}$ \\
\hline\hline
UA4/2 & 541 & 63.3 $\pm$ 1.5 \hs{0.5mm}$^{*}$ & 62.2 $\pm$ 1.5 & 0.135 $\pm$
 0.015 & 15.52 $\pm$ 0.07 \\
\hline
CDF & 546 & 62.64 $\pm$ 0.95 & 61.26 $\pm$ 0.93 & 0.15 \hs{0.5mm}$^{*}$ & 15.2
 $\pm$ 0.6 \\
\hline
CDF & 1800 & 81.83 $\pm$ 2.29 & 80.03 $\pm$ 2.24 & 0.15 \hs{0.5mm}$^{*}$ & 17.0
 $\pm$ 0.25 \\
\hline
E710 & 1800 & 73.6 $\pm$ 3.3 & 72.2 $\pm$ 2.7 & 0.134 $\pm$ 0.069 & 16.72 $\pm$
 0.44 \\
\hline
\end{tabular}
\end{center}

\hspace{10mm} Table 1. High-energy $\overline{p}p$ scattering parameters from
 most recent experiments. The assumed values are marked with
ASTERISKS(*).
 The E710 results are from the three parameter fits.
\normalsize

\vspace{5mm}
The experimental information about the scattering parameters for both $pp$ and
 $\overline{p}p$ scatterings are available only up to the ISR energies and at
 high energies only the $\overline{p}p $ scattering data are known  as given in
 Table 1.
The total cross sections started rising with the energy already at ISR
 and the $\overline{p}p$ cross
 section continues to increase up to the Tevatron energy.
However the difference between the CDF and E710 cross sections is posing a
 difficulty in predicting a unique asymptotic behavior of the cross sections as
 they both can be continued smoothly from the ISR data  and the
phenomenological models
 based on $\chi^2$ fit should favor neither of them as emphasized in
 [17] (See Figs. 1 and 2 in Ref.[17]). The absence of $pp$ cross
section data at
 higher energies is another reason why the asymptotic theorems are useful in
 model building.

The $t$-dependence of the differential cross sections away from the forward
 direction shows the diffraction pattern, i.e., a sharp diffraction peak
 followed by a diffractive minimum and secondary shoulder. The nuclear slope
 parameter $B$ is strictly speaking given by

\beq
B  = \frac{d}{dt} \ [ \hspace{1mm} \ln (\frac{d\sigma}{dt}) \hspace{1mm}
]_{t=0}
\eeq
and refers to the slope of the diffraction peak.
Experimentally $B$ \hs{1mm} shows a slow rise in energy which makes the
 diffraction peak to shrink as one can expect from a simple Regge behavior. A
 distinctive feature of $B$ at the ISR and also at the $Sp\overline{p}S$,
 though less pronounced, is the break in the slope around $\vert t \vert = 0.1
 \hspace{2mm}GeV^2$, which however seems to have almost disappeared at the
 Tevatron energy[18] ( See Figs. 4, 5 and 6 in Ref.[18] ).
At $\sqrt{s} = 53 \hspace{2mm}GeV $, the ISR data[19] show that the $pp$ and
 $\overline{p}p$ have almost identical diffraction structure with a diffractive
 minimum around $\vert t \vert = 1.3 \hspace{2mm} GeV^2 $ and secondary maximum
 at larger $\vert t \vert$-values except that there appears to be an extra
 shoulder filling in the $\overline{p}p$ minimum.
The $\overline{p}p$ data at the $Sp\ol{p}S$[20] shows almost no diffractive
 minimum but a broad shoulder which could be due to an experimental
resolution.
 The identity of the $pp$ and $\overline{p}p$ differential cross sections
within
 the diffractive minimum is expected from the asymptotic theorems as we will
see
 in the next Section but the question of difference around the diffractive
 minimum at high energies can not be answered from experiments for the
 indefinite future.
Apparent difference at the diffractive minimum between $pp$ and $\overline{p}p$
 at ISR and the rise of the secondary peak(shoulder) with energy for
 $\overline{p}p$ differential cross section prompted some to suggest the need
of
 three-gluon exchange or Odderon  beyond the Pomeron  and two-Pomeron
 cut (See Figs. 8 and 9 in Ref.[18]) as we mentioned before.

One of the most striking experimental features of the total cross sections for
 $pp$ and $\ol{p}p$ is that they have rather smooth behavior at energies above
a
 few $GeV $ up to the known Tevatron energy. This may suggest that the
 hadron-hadron scattering at high energies can be described by a simpler
 analyticity representation. Indeed a suggestion[10] was made
 in 1974 to employ
 certain quasi-local DDR to describe the high energy scattering data instead of
 the full dispersion relation which must be used at low energies to take into
 account the rich structure of total cross sections.

One can derive DDR from the Sommerfeld-Watson-Regge(SWR) representation
\beq
F_{\pm}(s,t) = - \frac{1}{2i} \ \int_{C} dj \hs{1mm} \frac{1 \pm e^{-i \pi
 j}}{\sin \pi j} \ s^j \hs{1mm} F(j,t)
\eeq
with the assumption that the asymptotic behavior is controlled by the
right-most
 singularity of $F(j,t) $ in the complex $j$-plane which in particular
is located near $j$  = 1 in forward scattering.
The SWR representation
LEADS
 to the quasi-local analytic relation in which the
 real and imaginary parts of $F_{\pm} $ are related to each other by certain
 derivatives with respect to $\ln s $. The leading terms of DDR at high
energies
 are

\beq
Re \hs{1mm}(F_{+}(s,t)\hs{1mm} / \hs{1mm}s) \simeq \frac{\pi}{2} \
 \frac{\partial}{\partial \ln s} \ Im \hs{1mm} (F_{+}(s, t)\hs{0.5mm} /
 \hs{0.5mm}s)
\eeq
\beq
Im \hs{1mm}(F_{-}(s,t)\hs{1mm} / \hs{1mm}s) \simeq  - \frac{\pi}{2} \
 \frac{\partial}{\partial \ln s} \ Re \hs{1mm} (F_{-}(s, t)\hs{0.5mm} /
 \hs{0.5mm}s)
\eeq
which reduce to the relationships between the total cross sections
 $\sigma_{\pm}(s) $ and real parts $\rho_{\pm}(s) $ in the forward direction.

It is easy to see from the SWR relation (13) that a simple Regge pole in the
 complex $j$-plane at $j = \alpha_k(t) $ gives
\beq
F_{+}^{(k)}(s,t) = C_{+}^{(k)}  [  \hs{1mm} i \hs{1mm} - \hs{1mm} \cot (
 \frac{\pi}{2} \ \alpha_k (t) ) ] \hs{1mm} s^{\alpha_k(t)}
\eeq
\beq
F_{-}^{(k)}(s,t) =  - \hs{1mm} C_{-}^{(k)} [ \hs{1mm} i \hs{1mm} + \hs{1mm
} \tan ( \frac{\pi}{2} \ \alpha_k (t) ) ] \hs{1mm} s^{\alpha_k(t)}
\eeq

The Regge pole $\alpha_k(t)$ is called exchange-degenerate
 if it contributes to
 both even- and odd-signatured amplitudes with $C_{+}^{(k)} = C_{-}^{(k)}$.
One can easily see that $F_{\pm}^{(k)}(s,t)$ are the solutions of the DDR (14)
 and(15) when $\alpha_k(t)$ is near 1. DDR possess the correct analyiticity
 property as given by the SWR representation.

The concept of the $Pomeron$ was introduced originally as a simple Regge
 pole at $j = \alpha_p(t) $ that has the quantum numbers of the vacuum and thus
 contributes to the even-signatured amplitude only so that both $F_{pp} $ and
 $F_{\ol{p}p} $ get the same contribution $F_{+}^{(P)} \sim s^{\alpha_p(t)} $
at
 high energies. As the cross sections can not increase faster than $\ln^2 s $
 from the rigorous results as we will see, $\alpha_{p}(0 )$ has to be very
close
 to 1. The slow and smoothly rising behavior of the total cross sections can
 then be approximated by a small power increase $s^{\epsilon_p} $, leading to
an empirical
 trajectory for the $Pomeron$
\beq
\alpha(t) = 1 + \epsilon_p + \alpha^{'} t = 1.08 + \alpha^{'} t
\eeq
The slope of the $Pomeron $ trajectory $\alpha^{'} $ can be determined from
 the slow increasing behavior of the nuclear slope parameter $B$ within the
 diffraction peak because the simple $Pomeron $ picture gives the explicit
 energy dependence of the slope parameter
\beq
 B = B_{o} + 2 \hs{1mm} \alpha^{'} \hs{1mm} \ln s
\eeq
and makes the diffraction peak to shrink as the energy increases.
Experimentally $\alpha^{'} = 0.2 GeV^{-2} $ as it has been known for about two
 decades [21]( See also Fig.2 in Ref.[18] ). The simple Regge pole picture of
 the Pomeron has a few other striking predictions. For example, the
 $\rho$-value for $pp$ as well as $\ol{p}p$ both should be given by
 $\frac{\pi}{2} \ \epsilon_p \simeq 0.12 $ at high energies and the elastic
 cross section $\sigma_{e}$ must increase faster than the total cross-section
 $\sigma_{T}$ because
\beq
 \sigma_e = \sigma_T^2 \hs{2mm} / 16 \pi B
\eeq
FOR
 the optical model type parametrization (8). Obviously  this effect can not
 continue forever.

There are in fact some authors[22] who do not want to be constrained by the
 asymptotic bounds and theorems at present energies and continue to prefer a
 simple
$bare~Pomeron$
picture. The need to unitarize the bare Pomeron
 behavior at high energies is however generally recognized, because the
 asymptotic bound is based on general principles involving more than just the
 unitarity and therefore
 less constraining. Thus one can violate the unitarity
 bound well below the rigorous asymptotic bound, which in fact is the case for
 the picture of the bare Pomeron [23]. Physically speaking, one may say
 the small but positive $\epsilon_p$ is
 a reflection of heavy flavour
production
 above a certain energy scale. Thus $\epsilon_p$ should be scale-dependent
 through new flavour production. One should then unitarize the bare Pomeron
 behavior by eikonalization.
If the bare Pomeron is due to the reggeized color-singlet two-gluon
 exchanges, one may say it should be renormalized through rescattering of two
 gluons via multi-gluon exchanges. This procedure will convert the bare Pomeron
 behavior to a more tamed $\ln ^{\gamma} s \hs{1.5mm}( \gamma \leq 2) $
 behavior.
The solutions we get from DDR corresponding to a singularity at $j = 1 $ in
THE
 forward direction correspond directly to the renormalized and physical
 Pomeron.

The solution of DDR that gives the asymptotic behavior of the total cross
 section to be $\ln^2 s $ is

\beq
F_{+}^{(P_2)}(s, 0) = i \hs{1mm} s \hs{1mm} [ A_{+} + B_{+}( \ln
\frac{s}{s_{+}}
 \ - i \frac{\pi}{2} \ )^2 ]
\eeq
which is sometimes called the unitarized or physical Pomeron term.

On the other hand, the odd-signatured counterpart
 of the Pomeron can
 also be constructed from DDR by assuming that the difference of the $pp$ and
 $\ol{p}p$ cross sections $\bigtriangleup\sigma_T = \sigma^{\ol{p}p}_T -
 \sigma^{pp}_T $ increases like $\ln s $ while the total cross sections
increase
 like $\ln^2 s $ [10, 24],

\beq
F_{-}^{(O)}(s, 0) = s \hs{1mm} [ A_{-} + B_{-}( \ln \frac{s}{s_{-}} \ - i
 \frac{\pi}{2} \ )^2 ]
\eeq
This odd-signatured Pomeron-like object is called
the maximal {\it Odderon.}
 One can also construct the Pomeron amplitude that gives the total
 cross section increasing with energy as $\ln s $,
\beq
F_{+}^{(P_1)}(s, 0) =  i \hs{1mm} s \hs{1mm} [ A_{+} + B_{+}( \ln
 \frac{s}{s_{+}} \ - i \frac{\pi}{2} \ ) ]
\eeq
In this case, $\bigtriangleup\sigma_T$ can be a non-zero constant
asymptotically,
\beq
F_{-}^{(O_1)}(s, 0) = s \hs{1mm} [ A_{-} + B_{-}( \ln \frac{s}{s_{-}} \ - i
 \frac{\pi}{2} \ ) ]
\eeq
Experimentally $\bigtriangleup\sigma_T \sim s^{\alpha(o) - 1}$ with $\alpha(0)
 \simeq 0.5 $ through the ISR energies.  It was shown from the eikonal
 formalism[13] that the maximal Odderon behavior is difficult to
 generate while either behavior of the two Pomeron amplitudes are easy
 and natural to derive.

In the
 eikonal formalism, the near-forward scattering amplitude for $pp$ and
 $\ol{p}p$ scattering is written as
\beq
F_{k}(s,t) = 4 \pi i s \int_{0}^{\infty} \hs{1mm} b \hs{1mm} db \hs{1mm}
 J_o(b\sqrt{-t}) \{ 1- e^{- \Omega_{k}(s,b)} \}, \hs{4mm} (k = pp, \hs{1mm}
 \ol{p}p)
\eeq
in the impact parameter $b$ space.
Here $ i \hs{1mm} \Omega_k(s,b)$ are the eikonals. One can then decompose the
 eikonals into $\Omega_{\pm} $ in analogous form as (2) so that
\beq
F_{+}(s,0) = 8 \pi i s \int_{0}^{\infty} \hs{1mm} b \hs{1mm} db \{ 1 - e^{ -
 \Omega_{+}(s,b)} \} \hs{2mm} cosh\hs{1mm} \Omega_{-}(s,b)
\eeq
\beq
F_{-}(s,0) = 8 \pi i s \int_{0}^{\infty} \hs{1mm} b \hs{1mm} db \hs{1mm} e^{ -
 \Omega_{+}(s,b)} \hs{2mm} sinh\hs{1mm} \Omega_{-}(s,b)
\eeq
There are several classes of the eikonals that can lead to $\ln^2 s $
 behavior for $\sigma_T$ but in no cases $\bigtriangleup\sigma_T \sim \ln s $
 behavior is possible to achieve.

Finally if $\sigma_T(s) \sim s^{\alpha_p(0) - 1}$ asymptotically, the
 Pomeron should also couple to the diffractive process $p + p \ra p + X $
 with $X$ having an invariant mass $M$ and diffractively produced, i.e.,
into the
 forward direction with the same quantum number as $p$.
Then through the unitarity  one will get the contributions from the diffractive
 process in the intermediate states through the optical theorem. Such
a process
 will produce the so-called $triple-Pomeron$ coupling contribution to the
 total cross section when both $s$ and $M^2$ are large compared to a
 characteristic scale beyond which the simple Regge picture makes sense. This
 process will obviously present a self-consistent check on the Pomeron
 dominance at high energies.
The CDF group[25] tested the standard triple-Pomeron Regge formula for
 single diffraction dissociation
\beq
s\hs{1mm} \frac{d^2\sigma_{SD}}{dt \hs{1mm} dM^2} \ = G(t) \hs{1mm}
 (\frac{s}{s_o} \ )^{\alpha_p(0)-1} \hs{1mm} (\frac{s}{M^2} \ )^{2\alpha_p(t) -
 \alpha_p(0)}
\eeq
at the Tevatron energies $\sqrt{s} = 546 \hs{1mm}GeV $ and $1800 \hs{1mm}GeV$
 for the regions $M^2/s < 0.2 $ and $0 \leq -t \leq 0.4 GeV^2 $.
They found in particular the need to include the screening corrections in the
 triple-Pomeron Regge model because the single diffraction total cross
 section shows a flat s-dependence rather than $s^{2\epsilon_p}$ as expected
 from the model.

The Pomeron exchange diagram in the diffraction scattering has an
 obvious similarity with the one in the deep-inelastic lepton scattering where
 the exchanged object is instead a virtual photon. Because of the similarity
 between the Pomeron and photon coupling to the quarks, there is a
 quantitative relation between the two processes and the deep-inelastic lepton
 scattering may be used to calculate the diffraction dissociation.
The recent experimental efforts at HERA are in fact to learn about, amongst
 others, the quark and gluon densities in the Pomeron, i.e., the
 partonic structure of the Pomeron based on the similarity with the
 diffraction dissociation processes [26].

\section{Asymptotic Theorems}

    As we mentioned before, experiments show that the nuclear slope
 parameter
 $B$ increases slowly with energy like $\ln s$ as expected from the simple
Regge
 Pomeron picture so that the elastic total cross section increases
 faster than the total cross section at the current energies.
For example, we have $\sigma_e/\sigma_T \simeq 0.18 $ at ISR energies $\sqrt{s}
 = 30$ to $60 \hs{1mm} GeV $; $0.215 \pm 0.005$ (UA4)[16] and $0.210 \pm 0.002$
 (CDF)[3] at $\sqrt{s} = 546 \hs{1mm}GeV$;  $0.23 \pm 0.012$ (E710)[27] and
 $0.248 \pm 0.005$ (CDF)[3] at $\sqrt{s} = 1800 \hs{1mm}GeV $.

Obviously the elastic total cross section can not continue to grow faster than
 the total cross section indefinitely so that from (20) the nuclear slope $B$
 should grow at least as fast as the total cross section at asymptotic energies
 within the framework of the optical model parametrization. In fact the
 rigorous statement is that the absorptive slope $B_A$ defined by the
absorptive
 part of the amplitude

\beq
B_{A}(s,t) = 2 \frac{d}{dt} \ ( \ln A(s,t) )
\eeq
satisfies the inequality
\beq
B_A(s,t=0) > \sigma_T^2/18 \pi \sigma_e^{A} > \sigma_T/ 18\pi
\eeq
a result which has been with us for three decades [28].
As for the full nuclear slope defined by (12), there can be complications
 because of the real part contributions and the best one can say[29] is that
 $B(s,t) < c(t) \ln s $ \hs{0.5mm}for $t < 0 $ under certain extra
assumptions of
 the infinite sequence of uniformly continuous functions.
While it is then clear that we are far from the asymptotic energy, the bound on
 the nuclear slope $B$ is an example indicating how general the rigorous
 statements are.

Nevertheless asymptotic theorems are very powerful and useful to guide what
 could be expected at high energies.
The most famous case of the rigorous statements is the Froissart bound
$\sigma_T
 < C \ln^2 s/s_{+}  $ with $C \leq \pi /m_{\pi}^2 \simeq 60 \hs{1mm} mb$ [30].
 Besides the unitarity condition for the partial waves, one needs to assume
 certain polynominal boundedness on the analytic scattering amplitude itself or
 its absorptive part to derive the Froissart bound.
As mentioned before, the Froissart bound is rather general and is not violated
 numerically for a reasonable choice of the scale parameter $s_{+}$ around a
few
 GeV$^2$ by the known experiments at current energies.
It is interesting to know if nature chooses to follow $\ln^2 s $ behavior for
 the total cross section at the asymptotic energy. The discrepancy between the
 CDF and E710 $\sigma_T $ at the Tevatron energy makes it difficult to draw an
 unambiguous conclusion in the asymptotic amplitude analysis as we will see in
 the next section.

There are several general statements that can follow when the Froissart bound
is
 saturated. One of them is the lower bound of the absorptive slope $B_A(s,t=0)$
 as discussed above. Another one is the Auberson, Kinoshita and Martin(AKM)
 theorem[31] which says if $\sigma_t \propto \ln^2 s $,
 $F_{\pm}(s,t)/F_{\pm}(s,0)$ tends to a non-trivial entire function of order
 $\frac{1}{2} \ $ of the variable $\tau = t \ln^2 s $ in the region $\vert t
 \vert < R / ln^2 s $. In this case, the diffraction peak shrinks like $1/
ln^2 s$.

With the increasing cross
 sections, the question of similarity between the
particle-particle and antiparticle-particle scattering cross sections,
i.e., the
 original Pomeranchuk theorem, $\sigma^{pp}_T - \sigma^{\ol{p}p}_T \ra 0 $
 as $s \ra \infty$, has to be replaced by the statement that
the ratio
 $\sigma^{\ol{p}p}_T / \sigma^{pp}_T$ tends to unity as $s \ra \infty$.
In fact,  more rigorous examinations revealed that the Pomeranchuk theorem
could be proven
 only when $\sigma_T^{pp}$ or $\sigma_T^{\ol{p}p}$ goes to infinity
as
 $s \ra \infty$[32]. In this case, the difference $\bigtriangleup\sigma_T $
 between $\sigma_T^{pp}$  and $\sigma_T^{\ol{p}p}$ does not necessarily tend to
 zero and the best one can say[33] is that $|\bigtriangleup\sigma_T| \leq C \ln
 s/s_{-} $ so that the maximal or other finite Odderon contributions are
 not contradicting with the rigorous statements at the asymptotic energy.

As for the amplitudes, if one allows all possible asymptotic behaviours
 for the
 even-signatured part consistent with the Froissart bound, i.e.,

\beq
F_{+}(s,t=0) \sim i \hs{1mm} B_{+} \hs{1mm} s \hs{1mm} [ \hs{1mm} \ln(s
\hs{1mm}
 e^{-i \pi/2} ) \hs{1mm} ]^{\beta_{+}}
\eeq
with $\beta_{+} \leq 2 $, the odd-signatured part can take any of the following
 form asymptotically [34],

\beq
F_{-}(s,t=0) \sim B_{-} \hs{1mm} s \hs{1mm} [ \hs{1mm} \ln(s \hs{1mm} e^{- i
 \pi/2} ) \hs{1mm} ]^{\beta_{-}}
\eeq
with $\beta_{-} \leq \beta_{+}/2 +1$ and $\beta_{-} < \beta_{+} + 1$.
In other words, there is a large domain of $(\beta_{+}, \beta_{-})$ that is
 allowed by the general principles of analyticity, unitarity and positivity of
 the absorptive part of $F_{pp}$ and $F_{\ol{p}p}$. The maximal Odderon
 behavior corresponds to the point $\beta_{+} = \beta_{-} = 2 $.

We note that experimentally $\sigma_{-} = \frac{1}{2} \ (\sigma_T^{pp} -
 \sigma_T^{\ol{p}p}) $ is negative up to the ISR energies but when the total
 cross sections increase with energy there is no guarantee that $\sigma_{-}$
 will continue to stay negative at higher energies in principle.
However $\sigma_{-}$ is consistent with $s^{-1/2}$ behavior from the known
 experiments.
If the odd-signature amplitude is negligible at high energies, one has
 $\rho_{pp} = \rho_{\ol{p}p}$ which will tend to $0$ as $s \ra \infty$.
But with $F_{-}$ contributions, one can have all sorts
 of possibilities for
 $\rho$. The rigorous statement of $\rho$ [35] based on general principles is
 that $\rho \leq \frac{\pi}{m_{\pi}} \ \frac{\ln (s/s_o)}{\sqrt{\sigma_T}} \ $.

The rigorous statement on the ratio of the differential cross sections for $pp$
 and $\ol{p}p$ at high energies is that inside the diffraction peak, the ratio
$(d\sigma / dt)_{\ol{p}p} / (d\sigma / dt)_{pp} $ tends to some limiting values
 that contain unity[36]. The proof of this statement involves several qualified
 assumptions such as those about the phase and imaginary parts of the
 amplitudes, But experimentally the ratio of the particle and antiparticle
 differential cross sections seems to approach to each other, i.e., the ISR
 data[19] at $\sqrt{s} = 52.8 \hs{1mm} GeV$ show remarkable equality except for
 the dip region as we mentioned above.

\section{Asymptotic Amplitude Analysis}

It is clear that the various asymptotic analytic representations for the
 Pomeron and Odderon discussed in Section II are all consistent with
 the rigorous asymptotic statements. The asymptotic amplitude analysis is then
 to construct the $pp$ and $\ol{p}p$ amplitudes from (3) where $F_{+}$ is made
 of various different combinations of the Pomeron $P_1$ or $P_2$ and
 Regge pole terms and $F_{-}$ given by Regge terms alone or plus
the Odderon term;
\beqa
F_{+} & = & F_{+}^{(P_i)}(s,0) + \sum_{k} F_{+}^{(k)}(s,0), \hs{15mm} (i = 1
 \hs{1mm}or \hs{1mm} 2)  \\
F_{-} & = & F_{-}^{(0)}(s,0) + \sum_{k} F_{-}^{(k)}(s,0), \hs{7mm}
or \hs{2mm} = \hs{2mm} \sum_{k} F_{-}^{(k)}(s,0)
\eeqa
which are then tested against the existing experimental data [11,17,37].

It is crucial then to select as complete a set of data as possible without
 leaving out any experimental group of data that  are published and not
 retracted, nor reducing errors to artificially enhance the weight to account
 for the paucity of higher energy results. Such a compilation of data,
including
 statistical merging of data points at a given energy, already exists[38].

Comprehensive $\chi^2$-fits and analysis have been carried out for two data
sets
 differing only by the lowest value of $\sqrt{s} $ allowed, i.e., 9.7
 \hs{1mm}$GeV$ and 5 $GeV$. There are 171 experimental points in the larger set
 of data for $\sqrt{s} \geq 5 \hs{1mm} GeV $ and 111 experimental points for
 $\sqrt{s} \geq 9.7 \hs{1mm} GeV $. They are distributed as follows: 97 values
 of $\sigma_T$ (40 for $\ol{p}p$ and 57 for $pp$), 65 values of $\rho$ (12 for
 $\ol{p}p$ and 53 for $pp$), and 9 values of $\bigtriangleup\sigma_T$ in the
 larger set, while 58 values of $\sigma_T$ (22 for $\ol{p}p$ and 36 for $pp$)
 and 53 values of $\rho$ (12 for $\ol{p}p$ and 41 for $pp$) in the smaller set.
 The details of the analysis can be found in [17]. Because the majority of
 precise data is at lower energies ( $\sqrt{s} < 63 \hs{1mm}GeV$ ), it is
 expected that somewhat detailed Regge terms beyond just one exchange
degenerate
 Regge pole term should be needed.
In addition because there is no new anomaly in higher energy experimental
 points, one can expect several forms of the combinations in $F_{+}$ and
$F_{-}$
 to do more or less equally well in terms of $\chi^2$.
In fact this was found to be the case for any model that contains more than
just
 one exchange-degenerate Regge term.
In particular,  Model $(B)$ in Ref. 17 in which $F_{+} = F_{+}^{(P_2)} +
 F_{+}^{R_d} + F_{+}^{R}$ and $F_{-} =  F_{-}^{R_d} + F_{-}^{R}$ where
 $F_{\pm}^{R_d}$ and $F{\pm}^R$  denote the exchange-degenerate and
non-degenerate
 Regge terms respectively, fits the data slightly better than  Model (D) of
Ref.
 17 in which $F_{+}^{(P_2)}$ is replaced by $F_{+}^{(P_1)}$ for $\sqrt{s} > 5
 \hs{0.5mm}GeV$ while the situation is reversed for the data set with $\sqrt{s}
 > 9.7 \hs{0.5mm}GeV$.
But the $\chi^2$-difference per d.o.f. in either case is only no larger than
 0.04, thus making both models equally acceptable in terms of statistics.
Note that in the first model, $\sigma_{+} \sim ln^2 s $ as $s \ra \infty$ while
 in the second model, $\sigma_{+} \sim ln s $ as $s \ra \infty$, thus  making
it
 difficult to favor either of the asymptotic forms of $\sigma_T$.
In either models however $F_{-} \ra 0$ as $s \ra \infty$  and in particular
 there is no need of the Odderon term in the asymptotic amplitude
 analysis.
In fact the maximal Odderon model fared better for the larger data set
 with $\sqrt{s} \geq 5 \hs{1mm}GeV$ than the data set with $\sqrt{s} \geq 9.7
 \hs{1mm} GeV$, implying that the Odderon term is effectively improving
 the low-energy fit and not relevant for the high-energy fit.

If the new CDF $\sigma_T$ is to withstand further experimental scrutiny, it may
 find a natural explanation in terms of a threshold slightly above the
 $Sp\ol{p}S$ energy. Detailed fit with  Model (G) in Ref.17 which is the
 modification of Model (D) in Ref.17 by a threshold term [39] gives $\sigma_T =
 77.1 \hs{1mm}mb$ at $\sqrt{s} = 1800 \hs{1mm} GeV $ somewhat smaller than the
 CDF $\sigma_T$ while the standard model (B) or (D) gives $\sigma_T = 74 \sim
76
 \hs{1mm}mb$ at the Tevatron energy which is consistent with the E710 value.
Since the UA4/2 data does not seem to require more than one exponential
 conforming to the standard expectation of $\rho$, the effective strength of
the
 threshold component, if it
 exists, should be rather weak leaving at most a few
 $mb$ level of new physics at the Tevatron energy. Finally Model (B) predicts
 $\sigma_T = 109 \hs{1mm} mb$ at the LHC while Model (D) gives $\sigma_T = 99
 \sim 104 \hs{1mm}mb$ compared to Model (G) which predicts $\sigma_T = 101 \sim
 105 \hs{1mm}mb$.

\vs{5mm} \large
{\bf Acknowledgement} \normalsize
\vs{5mm}

I would like to thank Professors D. P. Min and B. Y. Park for the stimulating
 environment and hospitality during the period of the Moojoo Summer School.

%\vs{20mm}
\newpage
\large
{\bf References} \normalsize
\vs{5mm}

{\bf 1.} UA4/2 Collab., C.Augier et al., Phys. Lett. {\bf B316}, 448(1993).

{\bf 2.} UA4 Collab., D. Bernard et al., Phys. Lett. {\bf B198}, 538(1987).

{\bf 3.} CDF Collab., F. Abe et al., Fermilab-Pub-93/234-E; See also

\hs{5mm} P. Giromini, in Proc. Vth Blois Workshop (Providence, RI,1993).

{\bf 4.} E710 Collab., S. Sadr, in Proc. Vth Blois Workshop; N. Amos et al.,

\hs{5mm} Phys. Lett. {\bf B243} 158(1990); Phys. Rev. Lett. {\bf 68},
 2433(1992).

{\bf 5.} E. M. Levin and M. G. Ryskin, Phys. Rep. {\bf 189}, No.6, 267(1990);

\hs{5mm} E. Levin, Fermilab-Pub-93/062-T.

{\bf 6.} L. Lipatov, in Perturbative QCD, ed. A. H. Mueller (World

\hs{5mm} Scientific, Singapore, 1989)

{\bf 7.} G. J. Daniell and D. A. Ross, Phys. Lett. {\bf B224}, 166(1989);

\hs{5mm} P. Gauron, L. Lipatov and B. Nicolescu, Phys.Lett. {\bf B260},
 407(1991).

{\bf 8.} L. V. Gribov, E. M. Levin and M. G. Ryskin, Phys. Rep. {\bf 100},

\hs{5mm} 1(1983); See also E. M. Levin (Ref. 5).

{\bf 9.} A. R. White, Mod. Phys. {\bf A6}, 1859(1991); also ANL-HEP-preprint.

{\bf 10.} K. Kang and B. Nicolescu, Phys. Rev. {\bf D11}, 2461(1974).

{\bf 11.} K. Kang, Nucl. Phys. B(Proc. Suppl.) {\bf 12}, 64(1990).

{\bf 12.} H. Cheng, J. K. Walker and T. T. Wu, Phys. Lett. {\bf 44B}, 97(1973);

\hs{6mm} C. Bourrely, J. Soffer and T. T. Wu, Phys. Rev. {\bf D19},3249(1979);

\hs{6mm} M. M. Block, in : Proc. Vth Blois Workshop where further related

\hs{6mm} references can be found.

{\bf 13.} J. Finkelstein, H. M. Fried, K. Kang and C-I Tan, Phys. Lett.

\hs{6mm} {\bf B232}, 257(1989).

{\bf 14.} A. Donnachie and P. V. Landshoff, Phys. Lett. {\bf 122B},345(1983);

\hs{6mm} Nucl. Phys. {\bf B231}, 189(1983); {\bf B244}, 322(1984).

\hs{6mm} See also P. Gauron, B. Nicolescu and E. Leader, Phys. Rev. Lett.

\hs{6mm} {\bf 54}, 2656(1985); Nucl. Phys.  {\bf B299}, 640(1988).

{\bf 15.} G. B. West and D. R. Yennie, Phys. Rev. {\bf 172}, 1413(1968);

\hs{6mm} N. H. Buttimore, E. Gostman and E. Leader, Phys. Rev. {\bf D18},

\hs{6mm} 694(1978); {\bf D35}, 407(1987); R. Cahn, Z. Phys. {\bf C15},
 253(1982).

{\bf 16.} U. Amaldi et al., Nucl. Phys. {\bf B145}, 367(1978);

\hs{6mm} UA4 Collab., M. Bozzo et al., Phys. Lett. {\bf 147B}, 392(1984).

{\bf 17.} K. Kang, P. Valin and A. R. White, in : Proc. Hadron '93

\hs{6mm}  (Como, Italy 1993) and Brown-HET-924.

{\bf 18.} M. M. Block, K. Kang and A. R. White, Mod. Phys. {\bf A7},

\hs{6mm} 444(1992), which reviewed the status of high energy elastic

\hs{6mm} scattering as of October 1991 before the UA4/2 and new CDF

\hs{6mm} results.

{\bf 19.} A. Breakstone et al., Phys. Rev. Lett. {\bf 54}, 2180(1985);

\hs{6mm} S. Erhan et al., Phys. Lett. {\bf B152}, 131(1985).

{\bf 20.} UA4 Collab., M. Bozzo et al., Phys. Lett. {\bf 155B}, 197(1985).

{\bf 21.} P. D. B. Collins, F. D. Gault and A. Martin, Nucl. Phys.

\hs{6mm} {\bf B85}, 141(1977).

{\bf 22.} A Donnachie and P. V. Landshoff, Nucl. Phys. {\bf B267}, 690(1986).

\hs{6mm} See also P. V. Landshoff, in : Proc. First Blois Workshop

\hs{6mm} (Ch\^ateau de Blois, France 1985); Proc. 3rd Blois Workshop

\hs{6mm} (Evanston, IL, 1989).

{\bf 23.} U. Maor, in : Proc. Vth Blois Workshop.

{\bf 24.} L. Lukaszuk and B. Nicolescu, Nuovo Cim. Lett. {\bf 8}, 405(1973).

{\bf 25.} CDF Collab., S. Belforte et al., CDF/ANAL/CDF/CDFR/2050.

\hs{6mm} See also P. Giromini in : Proc. Vth Blois Workshop.

{\bf 26.} G. Ingleman and P. Schline, Phys. Lett. {\bf B152}, 256(1985);

\hs{6mm} A. Donnachie and P. V. Landshoff, Nucl. Phys. {\bf B244} (Ref. 14)

{\bf 27.} N. Amos et al., Phys. Lett. {\bf B243} (Ref. 4)

{\bf 28.} S. W. Mac Dowell and A. Martin, Phys. Rev. {\bf 135}, 960(1964)

{\bf 29.} A. Martin, in : Proc. First Blois Workshop in which other

\hs{6mm} asymptotic bounds are also critically reviewed.

\hs{6mm} See also P. Valin, Phys. Rep. {\bf 203(4)}, 233(1991).

{\bf 30.} M. Froissart, Phys. Rev. {\bf 123}, 1053(1961);

\hs{6mm} A. Martin, Nuovo Cimento {\bf 42}, 930(1966);

\hs{6mm} L. Lukaszuk and A. Martin, Nuovo Cimento {\bf 47A}, 265(1967).

{\bf 31.} G. Auberson, T. Kinoshita and A. Martin, Phys. Rev.

\hs{6mm} {\bf D3}, 3185(1971).

{\bf 32.} T. Kinoshita, in : Perspectives in Modern Physics, ed. R. Marshak

\hs{6mm} (John Wiley and Sons, 1966); R, J. Eden, Phys. Rev. Lett. {\bf 16},

\hs{6mm} 39(1966); G. Grunberg and T. N. Truong, Phys. Rev. Lett. {\bf 31},

\hs{6mm} 63(1973).

{\bf 33.} S. M. Roy and V. Singh, Phys. Lett. {\bf 32B}, 50(1970);

\hs{6mm} R. J. Eden, Rev. Mod. Phys. {\bf 43}, 15(1971).

{\bf 34.} H. Cornille, Nuovo Cimento {\bf 70A}, 165(1970).

{\bf 35.} N. N. Khuri and T. Kinoshita, Phys. Rev. {\bf B140}, 706(1965).

{\bf 36.} H. Cornille and A. Martin, Nucl. Phys. {\bf B48}, 104(1972).

{\bf 37.} K. Kang, P. Valin and A. R. White, in: Proc. Vth Blois Workshop

\hs{6mm} and Brown-HET-916; M. M. Block et al., in : Proc. Muitiparticle

\hs{6mm} '93 Conference (Aspen, Co, 1993).

{\bf 38.} S. Hadjitheodoridis, Ph. D. thesis, Brown University(May 1989)

\hs{6mm}which is updated with high energy data by K. Kang, P.Valin and

\hs{6mm} A. R. White (Refs. 17 and 37).

{\bf 39.} K. Kang and S. Hadjitheodoridis, Phys. Lett. {\bf B208}, 135(1988).

\end{document}